\documentclass[10pt]{iopart}
\usepackage{bbm,iopams,mathrsfs,setstack}

\def\textbf#1{{\bf #1}}
\def\be{\begin{equation}}
\def\ee{\end{equation}}
\def\ben{\begin{eqnarray}}
\def\een{\end{eqnarray}}
\def\eea{\end{array}}
\def\bea{\begin{array}}
\newcommand{\ot}[0]{\otimes}
\newcommand{\bei}{\begin{itemize}}
\newcommand{\eei}{\end{itemize}}
\newcommand{\ket}[1]{|#1\rangle}
\newcommand{\bra}[1]{\langle#1|}

\newcommand{\proj}[1]{\ket{#1}\!\bra{#1}}

\newcommand{\mes}[0]{P_{m}^{+}}
\newcommand{\smes}[0]{\ket{\psi_{m}^{+}}}

\newcommand{\Trr}[0]{\mathrm{Tr}}

\def\blacksquare{\vrule height 4pt width 3pt depth2pt}

\begin{document}

\title[On structural physical approximations]{On structural physical approximations and entanglement breaking maps}

\author{R Augusiak$^{1}$, J Bae$^{2}$, \L{} Czekaj$^{3,4}$, and M Lewenstein$^{1,5}$}

\address{$^{1}$ICFO--Institut de Ci\`encies Fot\`oniques,
Parc Mediterrani de la Tecnologia, Castelldefels, 08860 Spain}
\address{$^{2}$School of Computational Sciences, Korea Institute for Advanced Study, Seoul 130--722, Korea}
\address{$^{3}$Faculty of Applied Physics and Mathematics,
Gda\'nsk University of Technology, Narutowicza 11/12, 80--952 Gda\'nsk,
Poland}
\address{$^{4}$National Quantum Information Centre of Gda\'nsk, Andersa 27, 81--824 Sopot, Poland}
\address{$^{5}$ICREA--Instituci\'o Catalana de Recerca i Estudis Avan\c{c}ats, Lluis Companys 23, 08010 Barcelona, Spain}

\begin{abstract}
Very recently a conjecture saying that the so-called structural
physical approximations (SPAa) to optimal positive maps (optimal
entanglement witnesses) give entanglement breaking (EB) maps
(separable states) has been posed [J. K. Korbicz {\it et al.},
Phys. Rev. A {\bf 78}, 062105 (2008)]. The main purpose of this
contribution is to explore this subject. First, we extend the set
of entanglement witnesses (EWs) supporting the conjecture. Then,
we ask if SPAs constructed from other than the depolarizing
channel maps also lead to EB maps and show that in general this is
not the case. On the other hand, we prove an interesting fact that
for any positive map $\Lambda$ there exists an EB channel $\Phi$
such that the SPA of $\Lambda$ constructed with the aid of $\Phi$
is again an EB channel. Finally, we ask similar questions in the
case of continuous variable systems. We provide a simple way of
construction of SPA and prove that in the case of the
transposition map it gives EB channel.
\end{abstract}

\maketitle

\section{Introduction}

Entanglement in composite quantum systems is a crucial resource in
the field of quantum information \cite{QIT,Horodecki09RMP}. It
allows to realize tasks which are not achievable in classical
physics with quantum teleportation \cite{Teleportation} being the
most prominent example.

It is thus natural that so much effort has been put so far in
developing various methods of detection of entanglement in
composite quantum systems \cite{GuhneToth09PR}. Among others,
probably the most powerful is the one based on the notion of
positive maps \cite{Peres96PRL,Horodecki96PLA,Horodecki01PLA}.
It states that a given state $\varrho$
acting on $M_{m}(\mathbbm{C})\ot M_{n}(\mathbbm{C})$ is separable,
i.e., it admits the decomposition \cite{Werner89PRA}:
\begin{equation}\label{separable}
\varrho=\sum_{i}p_{i}\proj{e_{i}}\ot\proj{f_{i}}, \qquad
\sum_{i}p_{i}=1,
\end{equation}
if and only if the Hermitian operator $(I\ot\Lambda)(\varrho)$ has
a nonnegative spectrum for any positive map
$\Lambda:M_n(\mathbbm{C})\to M_{m}(\mathbbm{C})$. Despite its
proven efficiency in entanglement detection, the above criterion
is not directly applicable in experiments. This is an immediate
consequence of the fact that generically positive, but not
completely positive (CP) maps do not represent physical processes.
It was then of a great importance to design methods which could
make the experimental detection of entanglement with the aid of
positive maps possible. So far, several such 
methods
(more or less direct) have been developed
\cite{HorodeckiEkert02PRL,MouraAlves03PRA,PHorodecki03PRA,Buscemi03PLA,Carteret05PRL,PHRAMD,RAMDPH}.

The earliest one, formulated and investigated in a series of
papers \cite{HorodeckiEkert02PRL,MouraAlves03PRA,PHorodecki03PRA},
is the so-called {\it structural physical approximation} (SPA). It
bases on the fact that any linear map (not necessarily positive)
when mixed with sufficiently large amount of the completely
depolarizing channel $D(\varrho)=\Tr(\varrho)\mathbbm{1}_{m}/m$,
which certainly is an element of the interior of the convex set of
CP maps, will result in a CP map. Since the latter already
represents some physical process, it can in principle be
implemented experimentally. Moreover, when applied to partial
maps, i.e., maps of the form $I\ot\Lambda$, the method produces CP
maps that hold the structure of the initial map in the sense that
one can easily recover the spectrum of $(I\ot\Lambda)(\varrho)$
knowing the spectrum of its SPA applied to $\varrho$. Let us
mention that a different approach to the approximation of the
transposition map by some CP map was presented in Ref.
\cite{Buscemi03PLA}. Under the assumption that such approximation
should work equally well for all pure states, the authors derived
the best CP approximation (in terms of fidelity) of the
transposition map. Interestingly, the obtained map matches SPA of
the transposition map. Eventually, it should be noticed that very
recently SPA of the transposition map was implemented
experimentally \cite{Bae10}.

It was observed for the first time by Fiur\'a\v{s}ek
\cite{Fiurasek02PRA} that SPA of the partial transposition map
$I\ot T$ in the two-qubit case is an entanglement breaking (EB) channel.
This is a very interesting observation meaning that the implementation of the SPA of
partial transposition reduces to generalized quantum measurement
followed by a preparation of a suitable quantum state conditioned
on the measurement outcomes \cite{HorodeckiShorRuskai03RMP}. Very
recently, this property was extensively investigated in Ref.
\cite{Korbicz08PRA}, where it was observed for many examples of
optimal positive maps\footnote{Notice that the authors
investigated SPAs of positive maps rather than the "extended" maps
$I\ot\Lambda$, which generally do not have to be positive.} that
their SPAs are entanglement breaking. Supported by the numerous
examples, the authors formulated the conjecture that
SPAs to optimal positive maps are
EB. It should be emphasized that in a series of papers
\cite{Pytel09,Pytel10} this conjecture was confirmed for new
classes of optimal entanglement witnesses (OEWs).

Due to its transparent relation to the geometry of the convex sets
of quantum states with positive partial transposition and sets of
entanglement witnesses \cite{Korbicz08PRA}, it is clear that
despite its physical meaning, the conjecture gains also in
importance from the mathematical point of view. It is then clear
that its complete proof is of a great interest. The main purpose
of the present paper is to address several issues related to this
subject.

The paper is organized as follows. In Sec. \ref{Preliminaries} all
the notions necessary for further studies are recalled. Sec.
\ref{SecIII} contains new examples of EWs supporting the
conjecture. Then, in Sec. \ref{Extending} we discuss several
issues related to the conjecture. In particular, we prove for any
positive map $\Lambda$ there exists an EB map $\Phi$ such that SPA
of $\Lambda$ constructed using $\Phi$ is EB. Also, we propose a
method of construction of SPAs in continuous variable systems, and
show that the SPA of the transposition map constructed with this
method gives EB channel. The paper is concluded with Sec.
\ref{Concl}.

\section{Background and the Conjecture}
\label{Preliminaries} 

First let us devote few lines to recall basic notions that will be
used frequently throughout the paper. In what follows, by
$\mathcal{H}_{mn}$, $\mathcal{H}_{m}$, and $M_{m,n}$ will be
denoting $\mathbbm{C}^{m}\ot\mathbbm{C}^{n}$,
$\mathbbm{C}^{m}\ot\mathbbm{C}^{m}$, $M_{m}(\mathbbm{C})\ot
M_{n}(\mathbbm{C})$, respectively. Then, $\mathcal{D}_{m,n}$
denote the set of all density matrices acting on
$\mathcal{H}_{m,n}$, i.e., positive matrices from $M_{m,n}$ with
unit trace. Finally, $\|\cdot\|_{2}$ will stand for the
Hilbert-Schmidt norm $\|A\|_{2}=\sqrt{\Tr(A^{\dagger}A)}$.

\subsection{Positive and completely positive linear maps and criterion for separability}

Following Ref. \cite{Werner89PRA} we call a state $\varrho$ acting
on $M_m(\mathbbm{C})\ot M_n(\mathbbm{C})$ {\it separable} if it
admits decomposition into a convex combination of products of
states describing local systems as in (\ref{separable}). The most
celebrated criterion allowing to distinguish separable states from
the entangled ones is the criterion based on positive maps
\cite{Peres96PRL,Horodecki96PLA} (see also Ref. \cite{Choi}).

We say that a linear map $\Lambda:M_m(\mathbbm{C})\to
M_n(\mathbbm{C})$ is {\it positive} if when acting on a positive
element from $M_m(\mathbbm{C})$, it returns positive element from
$M_n(\mathbbm{C})$. If, moreover, the extended map $I_d\ot\Lambda:
M_d(\mathbbm{C})\ot M_m(\mathbbm{C})\to M_d(\mathbbm{C})\ot
M_n(\mathbbm{C})$, where $I_d$ denotes identity map, is positive
for any $d$, we call $\Lambda$ {\it completely positive} (CP). A
linear map $\Lambda$ that do not change trace of the object on
which it acts, i.e., $\Tr[\Lambda(X)]=\Tr X$ is satisfied for any
$X\in M_m(\mathbbm{C})$, is called {\it trace-preserving} (TP).
Completely positive maps which are additionally TP are called {\it
quantum channels}. Let us mention that there is a simple criterion
allowing to judge if a given map is CP \cite{Horodecki99PRA}.
Namely, $\Lambda:M_m(\mathbbm{C})\to M_m(\mathbbm{C})$ is CP iff
$(I\ot\Lambda)(\mes)\geq 0$, where $\mes$ denotes a projector onto
the maximally entangled state
\begin{equation}
\smes=\frac{1}{\sqrt{m}}\sum_{i=0}^{m-1}\ket{i}\ket{i}.
\end{equation}

A well-known example of the positive, but not completely positive
map is the transposition map $T:M_m(\mathbb{C})\to
M_{m}(\mathbb{C})$. It is clear that when applied to a positive
operator $X$ it returns also a positive operator $X^{T}$. On the
other hand, local application of the transposition map (the
so-called partial transposition) to one of the subsystems of the
maximally entangled states gives operator $(I\ot T)(\mes)\equiv
(\mes)^{\Gamma}$ which is not positive. In what follows,  a
density matrix $\varrho$ such that $(I\ot T)(\varrho)\geq 0$
($(I\ot T)(\varrho)\ngeq 0$) will be called PPT (NPT).


A good example of a completely positive trace-preserving map is
the depolarizing channel $D[p]:M_m(\mathbbm{C})\to
M_m(\mathbbm{C})$ defined as
\begin{equation}
D[p](\varrho)=(1-p)\Tr(\varrho) \frac{\mathbbm{1}_{m}}{m}+p\varrho
\qquad (0\leq p\leq 1).
\end{equation}
When applied to a state $\varrho\in M_m(\mathbbm{C})$ it returns
$\varrho$ with probability $p$ and the maximally mixed state
$\mathbbm{1}_{m}/m$ with probability $1-p$. This map is CP because
its partial application to $\mes$ gives the known isotropic states
\cite{Horodecki99PRA}:
\begin{eqnarray}\label{iso}
\varrho_{\mathrm{iso}}(p)=\frac{1-p}{m^{2}}\mathbbm{1}_{m^{2}}+p\mes.
\end{eqnarray}
For brevity, we denote by $D$ the completely depolarizing channel
$D[1](\varrho)=\Tr(\varrho)\mathbbm{1}_{m}/m$.

The importance of positive maps in entanglement detection comes
from the fact that when acting locally on separable states, they
leave them in a separable form, and thus
$(I\ot\Lambda)(\varrho_{\mathrm{sep}})\geq 0$ for any separable
$\varrho_{\mathrm{sep}}$, and any positive map $\Lambda$. More
importantly, it was proven in Ref. \cite{Horodecki96PLA} that a
given state $\varrho$ acting on $\mathcal{H}_{m,n}$ is separable
if and only if the operator $(I\ot\Lambda)(\varrho)$ is positive
for any positive map $\Lambda:M_n(\mathbb{C})\to M_m(\mathbb{C})$.
This means that for any entangled state there exist a positive map
detecting it. Since, however, the structure of positive maps is
still not well understood, the sufficient part of this criterion
is not easy-to-apply. Nevertheless, it gives a very strong {\it
necessary criterion for separability}: given a state $\varrho$ and
a map $\Lambda$, check if the operator $(I\ot\Lambda)(\varrho)$
has negative eigenvalues, and if it does, then $\varrho$ is
entangled.

Within the set of positive maps let us distinguish the subset of
{\it decomposable maps} as the ones that admit the decomposition
$\Lambda=\Lambda_{1}^{\mathrm{CP}}+\Lambda_{2}^{\mathrm{CP}}\circ
T$ with $\Lambda_{i}^{\mathrm{CP}}:M_m(\mathbb{C})\to
M_n(\mathbb{C})$ denoting the completely positive maps, and $T$
the transposition map. All the remaining maps are called {\it
indecomposable}. Interestingly, since any positive map
$\Lambda:M_{n}(\mathbbm{C})\to M_2(\mathbbm{C})$  with $n=2,3$ is
decomposable \cite{Stormer,WoronowiczRMP76}, in the case of
qubit-qubit and qubit-qutrit systems we have a unique test
unambiguously detecting all entangled states. More precisely, all
the entangled states acting on $\mathbb{C}^{2}\ot\mathbb{C}^{m}$
with $m=2,3$ are detected by the transposition map.

Finally, among all the CP maps let us distinguish those which,
when partially applied to any bipartite state, give in return
separable states only. Such maps are called {\it entanglement
breaking} (EB), and if they are additionally TP, they are called
{\it entanglement breaking channels} (EBCs). These maps were
investigated in Ref. \cite{HorodeckiShorRuskai03RMP} (see also
Ref. \cite{Ruskai03RMP} and \cite{Holevo} for the characterization
of EB channels in continuous variables systems), and the
sufficient and necessary criterion allowing to check if a given
map is EB was worked out. Precisely, given CP map
$\Lambda:M_m(\mathbbm{C})\to M_m(\mathbbm{C})$ it is EB iff the
positive operator $(I\ot\Lambda)(\mes)$ can be written in a
separable form (\ref{separable}). Also, it was shown in Ref.
\cite{HorodeckiShorRuskai03RMP} that any entanglement breaking
channel can be represented in the Holevo form \cite{Holevo99RMS},
i.e., its action can be written as
\begin{eqnarray}\label{EB}
\Lambda(\varrho)=\sum_{k}\Tr(F_{k}\varrho)\varrho_{k},
\end{eqnarray}
where the operators $F_k\in M_m(\mathbbm{C})$ acting on represent
some generalized quantum measurement, i.e., they are positive and
sum up to $\mathbbm{1}_{m}$, and $\varrho_{k}$ stand for some
density matrices acting on $\mathcal{H}_m$.

\subsection{Entanglement witnesses and their optimality}
\label{optimality}

From experimental point of view it is important to notice that the
above separability criterion can be reformulated in terms of mean
values of some physical observables \cite{Horodecki96PLA}. To be
more precise let us recall that the Choi-Jamio\l{}kowski (C-J)
isomorphism \cite{Jamiolkowski,Choi75LAA} assigns to every
positive map $\Lambda:M_n(\mathbb{C})\to M_{n'}(\mathbb{C})$, a
Hermitian operator
\begin{eqnarray}\label{witnesses}
W_\Lambda=(I\ot\Lambda^{\dagger})(P_{n'}^{+}),
\end{eqnarray}
acting on $M_{n'}(\mathbbm{C})\ot M_n(\mathbbm{C})$ with
$\Lambda^{\dagger}:M_{n'}(\mathbbm{C})\to M_{n}(\mathbbm{C})$
denoting the dual map of $\Lambda$.

Since mean values of operators given by Eq. (\ref{witnesses}) on
separable states are always nonnegative, and, on the other hand,
for any $W$ there exists an entangled state $\rho$ for which
$\langle W\rangle_{\rho}<0$, we call these operators {\it
entanglement witnesses} \cite{Terhal00PLA}.

Via the C-J isomorphism, the division of positive maps to
decomposable and indecomposable can be translated to EWs. More
specifically, the decomposable positive maps give {\it
decomposable entanglement witnesses} which, as it follows directly
from Eq. (\ref{witnesses}), take the general form $W=P+Q^{\Gamma}$
with $P,Q\geq 0$. If a given witness cannot be written in this
form we call it {\it indecomposable}. They can be written as
$W_{\mathrm{in}}=P+Q^{\Gamma}-\epsilon\mathbbm{1}$ with
$0<\epsilon\leq \inf_{\ket{e,f}}\bra{e,f}P+Q^{\Gamma}\ket{e,f}$
and as previously $P,Q\geq 0$ and there exists some PPT edge
state\footnote{We call $\delta\in M_m(\mathbbm{C})\ot
M_n(\mathbbm{C})$ {\it a PPT edge state} if it is entangled PPT
state and for any product vector $\ket{e,f}\in\mathcal{H}_{mn}$
and $\epsilon>0$ the operator $\delta-\epsilon\proj{e,f}$ either
is not positive or it has nonpositive partial transposition.}
\cite{LewensteinSanpera98PRL} such that its kernel is equal to the
support of $P$, while kernel of its partial transposition to the
support of $Q$ \cite{LewensteinKraus01PRA} (see also Ref.
\cite{Bruss02JMO}).

Of all the EWs the ones which merit distinction are the {\it
optimal} ones. Precisely, following Ref.
\cite{LewensteinKraus00PRA}, we call $W$ {\it finer } than $W'$ if
the set of states detected by $W$ (denoted $D_W$) contains the set
of states detected by $W'$, i.e., $D_W'\subseteq D_W$. The witness
$W$ is called optimal if there does not exist a witness finer than
$W$. Let us briefly remind some relevant facts about optimal
witnesses obtained in Ref. \cite{LewensteinKraus00PRA}. The above
definition means that $W$ is optimal if and only if
$(1+\epsilon)W-\epsilon P$ is no longer an EW for any $\epsilon>0$
and $P\geq 0$. Then, if the set
$\mathcal{P}_{W}=\{\ket{e,f}|\bra{e,f}W\ket{e,f}=0\}$ spans the
Hilbert space on which $W$ acts then it is optimal. These facts
immediately imply that decomposable witnesses are only those which
take the form $W=Q^{\Gamma}$ with positive $Q$ which does not
contain any product vectors in its support.

Restricting to indecomposable witnesses only, the above properties
have to be reformulated. An operator $W$ is indecomposable optimal
entanglement witness iff for any $\epsilon>0$ and $P,Q\geq 0$, the
operator $(1+\epsilon)W-\epsilon(P+Q^{\Gamma})$ is not an
entanglement witness (one cannot subtract a decomposable
operator). Moreover, if both $\mathcal{P}_{W}$ and
$\mathcal{P}_{W^{\Gamma}}$ span the whole Hilbert space, then $W$
is indecomposable optimal entanglement witness.

One could conjecture that the above conditions for optimality of
witnesses is also sufficient. There exists, however, an example of
an indecomposable map provided by Choi \cite{ChoiMap} (see Sec.
\ref{ClassIn}) such that its $\mathcal{P}_W$ does not span the
corresponding Hilbert space, but it leads to a optimal
entanglement witness. The latter follows from the fact that the
Choi map was proven to be extremal in the set of positive maps
\cite{ChoiLam}.

\subsection{Structural physical approximation and the conjecture}
\label{SPAConj}

The C-J isomorphism tells us that both formulations of the above
separability criterion are equivalent. Nevertheless, a given
positive map detects more states that the corresponding
entanglement witness. The problem with the formulation of the
criterion in terms of positive maps lies in the fact that
generically they do not represent any physical process.
Consequently, they cannot be implemented experimentally. Since,
they are more efficient in entanglement detection, it has been an
interesting question how to make positive maps experimentally
implementable.

One way to solve this problem employs the {\it structural physical
approximation} (SPA)
\cite{HorodeckiEkert02PRL,MouraAlves03PRA,PHorodecki03PRA}.
Roughly speaking, the idea behind the SPA comes from the fact that
any positive, but not completely positive map $\Lambda$ (actually
it does not have to be even positive), when mixed with
sufficiently large amount of the completely depolarizing channel
$D$, will result in CP map. Precisely, the idea is to take the
following mixture
\begin{eqnarray}\label{FinSPA}
\widetilde{\Lambda}[p]=(1-p)D+p\Lambda \qquad (0\leq p\leq 1).
\end{eqnarray}
From the C-J isomorphism it is clear that there exists a
nontrivial parameter range $0\leq p\leq p_{*}$ with $p_*>0$ for
which the map $\widetilde{\Lambda}[p]$ is completely positive
(represents quantum channel if the initial map $\Lambda$ is TP),
and thus, in principle, it can represent some physical process.
The least noisy completely positive map from the class
$\widetilde{\Lambda}[p]$ $(0\leq p\leq p_{*})$, i.e.,
$\widetilde{\Lambda}[p_{*}]$ is called structural physical
approximation of $\Lambda$.

In the language of entanglement witnesses SPA is a positive
operator obtained by mixing a given witness $W$ from $M_{m,n}$
with maximally mixed state $\mathbbm{1}_{m}\ot\mathbbm{1}_{n}/mn$,
i.e.,
\begin{eqnarray}\label{SPAWit}
\widetilde{W}(p)=(1-p)\frac{\mathbbm{1}_{m}\ot\mathbbm{1}_{n}}{mn}+pW\qquad
(0\leq p\leq 1).
\end{eqnarray}
Since the maximally mixed state is represented by a full rank
positive matrix it is clear that there exists a threshold value
for $p$ (denoted by $p_*$) for which $\widetilde{W}(p)$ becomes
positive. Simple calculations show that it is given by
$p_*=1/(1+mn\lambda)$, where $-\lambda<0$ is the smallest
eigenvalue of $W$.

Having all the necessary notions and tools we can recall the
conjecture formulated in Ref. \cite{Korbicz08PRA}.\\

\noindent{\bf Conjecture.} {\it Let $\Lambda$ be an optimal
(trace-preserving) positive map. Then, its SPA is entanglement
breaking map (channel).\\}

The difficulty of proving or disproving the conjecture comes from
whether the density matrices resulting from an application of SPA
to EWs are separable, and the separability problem still remains
unsolved.

In what follows, basing on what was said in Sec. \ref{optimality},
we prove two facts relating the notions of SPA and optimality
of EWs.\\

\noindent{\bf Fact 1.} {\it Let $W_1$ and $W_2$ be two entanglement
witnesses. If $W_1$ is finer than $W_2$ then $p_*^{(1)}\leq
p_*^{(2)}$, where $p_*^{(1)}$ and $p_*^{(2)}$ are threshold values
in (\ref{SPAWit}) of $W_1$ and $W_2$, respectively. In this sense
optimal witnesses have smallest $p_*$.\\}

\noindent{\bf Proof.} Let $D_{W_1}$ and $D_{W_2}$ be the sets of states
detected by $W_1$ and $W_2$, respectively. Since $W_1$ is finer
than $W_2$, $D_{W_2}\subseteq D_{W_1}$. Let also
$-\lambda_{1(2)}<0$ denote the smallest eigenvalue of $W_1$
($W_2$). It is clear that
\begin{equation}
-\lambda_{1,2}=\min_{\rho\in D_{W_{1,2}}}\Tr(W_{1,2}\rho).
\end{equation}
On the other hand, we already know that
$p_{*}^{(1,2)}=1/(1+mn\lambda_{1,2})$. Then, in order to prove
that $p_{*}^{(1)}\leq p_{*}^{(2)}$, it suffices to apply the
following estimation
\begin{eqnarray}
-\lambda_{1}&=&\min_{\rho\in
D_{W_{1}}}\Tr(W_{1}\rho)\nonumber\\
&\leq& \min_{\rho\in D_{W_{2}}}\Tr(W_{1}\rho)\nonumber\\
&\leq& \min_{\rho\in D_{W_{2}}}\Tr(W_{2}\rho)=-\lambda_2,
\end{eqnarray}
where the second inequality is a consequence of the fact that
$W_1$ is finer than $W_2$ and therefore $\Tr(W_1\rho)\leq
\Tr(W_2\rho)$ for all $\rho\in D_{W_2}$. $\blacksquare$\\

\noindent{\bf Fact 2.} {\it Suppose that $W_1$ is finer than $W_2$. Then,
the SPA of $W_1$, $\widetilde{W}_1$, cannot be detected by
$W_2$.\\}

\noindent{\bf Proof.} Since $W_1$ is finer than $W_2$, one can
express the latter as $W_2 = (1-\epsilon)W_1 + \epsilon P$ for
some $\epsilon > 0$ and $P\geq0$. Since
\begin{equation}
\Trr(\widetilde{W}_1W_1)=p_{*}^{(1)} \Tr W^{2}_1 +
[(1-p_{*}^{(1)})/m^{2}]\Tr W_1 \geq 0,
\end{equation}
and $\widetilde{W}_1\geq 0$, it is straightforward to see
\be \Trr(\widetilde{W}_1 W_2) = (1-\epsilon)
\Trr(\widetilde{W}_1W_1) + \epsilon \Trr(\widetilde{W}_1P)\geq 0,
\ee
which completes the proof. $\blacksquare$

This fact shows that any non-optimal EW $W_2$ that can be
optimized to an optimal EW $W_1$ cannot detect the SPA
$\widetilde{W}_1$. Hence, provided $\widetilde{W}_1$ is entangled,
the EW detecting it cannot be the one that can be optimized to
$W_1$. This, somehow, reflects the difficulty of proving or
disproving the conjecture.

\section{Extending the set of witnesses which support the conjecture}
\label{SecIII}

The main aim of this section is to provide new examples of optimal
entanglement witnesses which obey the above conjecture.

\subsection{All decomposable witnesses with $Q$ of rank one}

Let us start from the decomposable witnesses being partial
transpositions of pure entangled states.\\

\noindent{\bf Theorem 1.} {\it Let $W=\proj{\psi}^{\Gamma}$ be a
witness with entangled $\ket{\psi}\in\mathcal{H}_m$. Then its SPA
is entanglement breaking channel.\\}

\noindent{\bf Proof.} First we write $\ket{\psi}$ in its Schmidt
decomposition:
\begin{equation}
\ket{\psi}=\sum_{i=0}^{r-1}\mu_{i}\ket{ii},
\end{equation}
where $r>1$, and $\mu_i\geq 0$ denote the Schmidt rank and Schmidt
coefficients of $\ket{\psi}$, respectively. Also, without any loss
of generality, we can set Schmidt local bases to the computational
ones. Following Eq. (\ref{SPAWit}), the structural physical
approximation of $W$ reads
\begin{eqnarray}\label{SPAproj}
\widetilde{W}(p)=(1-p)\frac{\mathbbm{1}_{m}\ot\mathbbm{1}_{m}}{m^{2}}+p\proj{\psi}^{\Gamma}.
\end{eqnarray}
It is clear that the eigenvalues of $\proj{\psi}^{\Gamma}$ are
$\mu_{i}^{2}$ $(i=1,\ldots,r)$ and $\pm\mu_{i}\mu_{j}$ $(i\neq
j=1,\ldots,r)$. Therefore, denoting by $\theta$ the largest of the
eigenvalues $\mu_{i}\mu_{j}$ $(i\neq j)$, we see that
$\widetilde{W}(p)\geq 0$ iff $p\leq 1/(d^{2}\theta+1)$. Now, our
aim is to prove that for $p_{*}=1/(d^{2}\theta+1)$, the state
$\widetilde{W}(p_{*})$ is separable. For this purpose, let us
consider the following class of states
\begin{eqnarray}\label{class}
\rho(p)\equiv
[\widetilde{W}(p)]^{\Gamma}=(1-p)\frac{\mathbbm{1}_{m}\ot\mathbbm{1}_{m}}{m^{2}}+p\proj{\psi}.
\end{eqnarray}
If the state $\rho(p)$ is separable for $p=p_{*}$ then obviously
$[\rho(p_{*})]^{\Gamma}=\widetilde{W}(p_{*})$ also is. For the
threshold value $p=p_{*}$, the above reduces to
$\rho(p_{*})=[1/(d^{2}\theta+1)]\left(\theta\mathbbm{1}_{m}\ot\mathbbm{1}_{m}+\proj{\psi}\right)$,
which after simple movements can be rewritten as
\begin{eqnarray}
\rho(p_*)&=&
\frac{1}{d^{2}\theta+1}\left(\theta\sum_{i=0}^{r-1}\proj{ii}+
\sum_{\underset{i\neq j}{i,j=0}}^{r-1}(\theta-\mu_{i}\mu_{j})\proj{ij}+\theta\sum_{i,j=r}^{m-1}\proj{ij}\right.\nonumber\\
&&\hspace{1.5cm}\left.+
\sum_{\underset{i\neq
j}{i,j=0}}^{r-1}\mu_{i}\mu_{j}\proj{ij}+\sum_{i,j=0}^{r-1}\mu_{i}\mu_{j}\ket{ii}\!\bra{jj}\right).
\end{eqnarray}
%

Due to the fact that $\theta\geq \mu_i\mu_j$ for any $i<j$
$(i,j=0,\ldots,r-1)$, one sees that first three terms represent
separable states. Consequently, what remains is to prove the
separability of the last two terms. It is clear that they give a
positive matrix, and thus represent a (unnormalized) quantum
state. After normalization we can rewrite them as
\begin{equation}
\overline{\rho}_{\mu}=\frac{1}{N}\left( D_{\mu}\ot
D_{\mu}+\sum_{i\neq j}\mu_{i}\mu_{j}\ket{ii}\!\bra{jj}\right),
\end{equation}
where by $D_{\mu}$ we denote a matrix diagonal in the standard
basis with eigenvalues $\mu_{i}$ $(i=0,\ldots,r-1)$, while
$N=(\Trr D_{\mu})^{2}$. Separability of this state for any
configuration of $\mu$ was proven in Ref. \cite{VidalTarrach}.
Consequently, we see that $\widetilde{W}(p_{*})$ can be written as
a convex combination of separable states and therefore is itself
separable. $\blacksquare$

Let us remark that, since every witness $W=\proj{\psi}^{\Gamma}$
with entangled $\ket{\psi}$ is extremal in the convex set of
entanglement witnesses, it is also optimal. Also, notice that the
above theorem can be extended easily to the case of unequal
dimension on both sites.

\subsection{A class of indecomposable positive maps}
\label{ClassIn}

In Ref. \cite{Korbicz08PRA} the conjecture was proven for the Choi
map \cite{ChoiMap}. Here, we show that this analysis can be
extended to one of the generalizations of Choi map analyzed in a
series of papers
\cite{TanahashiTomiyama88CMB,Osaka93LAA,Ha98PRIMS}. Their action
on any $X\in M_m(\mathbbm{C})$ can be expressed as
\begin{equation}
\tau_{m,k}(X)=(m-k)\varepsilon(X)+\sum_{i=1}^{k}\varepsilon(S^{i}XS^{i\dagger})-X,
\end{equation}
where $k=1,\ldots,m-1$ and the completely positive map
$\varepsilon$ is given by
$\varepsilon(X)=\sum_{i=0}^{m-1}\bra{i}X\ket{i}\proj{i}$. Note
that $k=m-1$ the map $\tau_{m,m-1}$ is just the reduction map
$R_{-}(X)=\Tr(X)\mathbbm{1}_{m}-X$
\cite{Horodecki99PRA,Cerf99PRA}. For the remaining cases,
$k=1,\ldots,m-2$, the map is indecomposable, and in particular
$\tau_{3,1}$ is the Choi map. Eventually, $S$ is a shift operator
acting as $S\ket{i}=\ket{i+1}$ (mod $m$).

Now, let us introduce the normalized version of the above map,
that is $\mathcal{T}_{m,k}=[1/(m-1)]\tau_{m,k}$, and notice that
it is TP and unital\footnote{We call the map
$\Lambda:M_m(\mathbb{C})\to M_m(\mathbb{C})$ unital if
$\Lambda(\mathbbm{1}_{m})=\mathbbm{1}_{m}$.}. Acting with the
extended map $I\ot \mathcal{T}_{m,k}$ on the maximally entangled
state, one obtains
\begin{equation}
W_{m,k}=\frac{1}{m(m-1)}\left[
(m-k)\varrho_{0}+\sum_{i=1}^{k}\varrho_{i} -m\mes\right]
\end{equation}
with $\varrho_{i}$ $(i=0,\ldots,k)$ denoting (unnormalized)
separable density matrices of the form
\begin{eqnarray}\label{SepStates}
\varrho_{i}=\sum_{l=0}^{m-1}\proj{l}\ot\proj{l+i},
\end{eqnarray}
where the summation is modulo $m$. By mixing $W_{m,k}$ with the
maximally mixed state, we arrive at
\begin{equation}
\widetilde{W}_{m,k}(p)=\frac{1-p}{m^{2}}\mathbbm{1}_{m^{2}}+pW_{m,k},
\end{equation}
which, due to the fact that the smallest eigenvalue of $W_{m,k}$
is $-k/[m(m-1)]$, is positive iff $p\leq [(m-1)/(m(k+1)-1)]$. Now,
we need to show that for $p_{*}=(m-1)/[m(k+1)-1]$, the matrix
$\widetilde{W}_{m,k}(p_{*})$ represents a separable state. Let us
then rewrite it as
\begin{equation}\label{rownanie}
\widetilde{W}_{m,k}(p_{*})=\frac{1}{N}
\left[(m-k)\varrho_{0}+k\mathbbm{1}_{m^{2}}-m\mes+\sum_{i=1}^{k}\varrho_{i}\right],
\end{equation}
where $N=m[m(k+1)-1]$. As the last term in the square brackets in
the above is already separable (cf. (\ref{SepStates})), our aim is
to prove the separability of the first three terms. For this
purpose, we can follow the reasoning applied to the Choi map in
Ref. \cite{Korbicz08PRA}. More precisely, we can rewrite these
terms as
\begin{equation}\label{decomposition}
(m-k)\varrho_{0}+k\mathbbm{1}_{m^{2}}-m\mes=
\sum_{\underset{i<j}{i,j=0}}^{m-1}\sigma_{ij}+(k-1)(\mathbbm{1}_{m^{2}}-\varrho_{0}),
\end{equation}
where
$\sigma_{ij}=\proj{ii}+\proj{jj}+\proj{ij}+\proj{ji}-\ket{ii}\!\bra{jj}-\ket{jj}\!\bra{ii}$
stand for two--qubit matrices embedded in
$M_{m^{2}}(\mathbbm{C})$. One easily checks that they are PPT and
thus separable. The second term, being diagonal in the standard
basis in $\mathcal{H}_m$, is also separable. Consequently,
$\widetilde{W}_{m,k}(p_{*})$ represents a separable state for all
possible values of $m$ and $k$, meaning the SPAs of
$\mathcal{T}_{m,k}$ are EB channels.

Concluding, let us comment on the optimality of the witnesses
$W_{m,k}$. In Ref. \cite{Korbicz08PRA} it was shown that in the
case of the Choi map, $\tau_{3,1}$, the set
$\mathcal{P}_{W_{3,1}}$ do not span the corresponding Hilbert
space $\mathbbm{C}^{3}\ot\mathbbm{C}^{3}$. Nevertheless, since
$\tau_{3,1}$ is extremal, it has to be optimal as one can subtract
neither a decomposable witness nor a positive operator. The
reasoning from Ref. \cite{Korbicz08PRA} can be extended to all
maps $\tau_{m,k}$ $(k=1,\ldots,m-1)$. Specifically, one checks
directly that the product vectors from $\mathcal{P}_{W_{m,k}}$ can
be written as
\begin{equation}
\sum_{k=0}^{m-1}\mathrm{e}^{\mathrm{i}\phi_{k}}\ket{k}\ot
\sum_{k=0}^{m-1}\mathrm{e}^{-\mathrm{i}\phi_{k}}\ket{k}\qquad
(\phi_{0}=0).
\end{equation}
Clearly, they span $m^{2}-m+1$ dimensional subspace of
$\mathcal{H}_m$, and therefore the problem of optimality of these
witnesses (except for $W_{3,1}$) is open.

\subsection{The conjecture and the isotropic states}

Very recently a simple, but useful fact relating the conjecture,
and the isotropic states (\ref{iso})
was proven \cite{Acin,Pytel09}. It can be stated as follows. \\

\noindent {\bf Theorem 2 \cite{Acin,Pytel09}.} {\it Let
$\Lambda:M_m(\mathbb{C})\to M_m(\mathbb{C})$ be a unital map that
detects all the entangled isotropic states
$\varrho_{\mathrm{iso}}(p)$ (cf. (\ref{iso})).
Then SPA of $\Lambda$ is an entanglement breaking map.} \\

The aim of this section is two-fold. On the one hand, we determine
the class of decomposable positive maps which satisfy assumptions
of this theorem and therefore obey the conjecture. On the other
hand, we provide a simple generalization of this theorem to the
class of maps containing, among others, unital maps.

In order to look for the witnesses satisfying theorem 2, let us
rephrase is slightly. Denoting by $\Lambda$ some unital map, the
condition of detecting all the entangled isotropic states by this
map is equivalent to the condition that there exists a pure
entangled state $\ket{\psi}\in\mathcal{H}_m$ such that
\begin{eqnarray}\label{c0}
\Tr[(I\ot\Lambda^{\dagger})(P_{\psi})\varrho_{\mathrm{iso}}(\textstyle\frac{1}{d+1})]
=\Tr[(A\ot\mathbbm{1}_{m})\omega(A^{\dagger}\ot\mathbbm{1}_m)
\varrho_{\mathrm{iso}}(\textstyle\frac{1}{d+1})]=0.\nonumber\\
\end{eqnarray}
Here $P_{\psi}=\proj{\psi}$,
$\omega=(I\ot\Lambda^{\dagger})(\mes)$ denotes the witness
corresponding to $\Lambda$ via the C-J isomorphism, and $A$ stands
for the matrix such that $\ket{\psi}=A\ot\mathbbm{1}_{m}\smes$.
Notice that since $\Lambda$ is unital, its dual map
$\Lambda^{\dagger}$ is TP. It is now clear that our aim is to find
the general form of decomposable witnesses $\omega$ satisfying the
above condition.

Since, by assumption, $\Lambda$ is decomposable,
$\omega=P+Q^{\Gamma}$ with $P,Q\geq0$ (see Sec. \ref{optimality}).
Also, we know that $\Lambda$ is unital if and only if the
corresponding EW $\omega$ has a maximally mixed first subsystem
$(\Trr_B\omega=\mathbbm{1}_m/m)$ meaning that $P$ and $Q$ have to
obey $\Trr_B(P+Q)=\mathbbm{1}_{m}/m$.

Denoting
$P_A=(A\ot\mathbbm{1}_{m})P(A^{\dagger}\ot\mathbbm{1}_{m})$ (the
same for $Q$), we can now rewrite the condition (\ref{c0}) as
\begin{equation}
\Tr[P_{A}\varrho_{\mathrm{iso}}(\textstyle\frac{1}{d+1})]
+\Tr[Q_{A}\varrho^{\Gamma}_{\mathrm{iso}}(\textstyle\frac{1}{d+1})]=0.
\end{equation}
We know that $\varrho_{\mathrm{iso}}(\textstyle\frac{1}{d+1})$ is
separable and therefore PPT. This together with positivity of
$P_A$ and $Q_A$ allows us to infer that both terms in the above
equation have to be zero, i.e.,
\begin{eqnarray}\label{cond1}
\Tr[P_{A}\varrho_{\mathrm{iso}}(\textstyle\frac{1}{d+1})]=0,\qquad
\Tr[Q_{A}\varrho^{\Gamma}_{\mathrm{iso}}(\textstyle\frac{1}{d+1})]=0.
\end{eqnarray}
From the explicit form of
$\varrho_{\mathrm{iso}}(\textstyle\frac{1}{d+1})=[1/d(d+1)](\mathbbm{1}_{m^{2}}+dP_{+}^{(m)})$
one sees that it is of full rank and hence the first condition in
Eq. (\ref{cond1}) yields $P_{A}=0$. On the other hand, one checks
that
$\varrho^{\Gamma}_{\mathrm{iso}}(\textstyle\frac{1}{d+1})=[2/d(d+1)]P_{\mathrm{s}}$
with $P_{\mathrm{s}}$ denoting the projector onto the symmetric
subspace of $\mathcal{H}_m$ (henceforward denoted
$\mathcal{S}(\mathcal{H}_m)$ or shortly $\mathcal{S}_m$). This,
when applied to the second condition in Eq. (\ref{cond1}), implies
that $Q_{A}$ has to be supported\footnote{By saying that a
positive operator $Q$ is supported on a subspace $V$ we mean that
its eigenstates corresponding to nonzero eigenvalues belong to
$V$. } on the antisymmetric subspace of $\mathcal{H}_m$
(henceforward denoted $\mathcal{A}(\mathcal{H}_m)$ or shortly
$\mathcal{A}_{m}$).

Let us remind that the antisymmetric subspace of $\mathcal{H}_m$
is spanned by the vectors
$\ket{\varphi_{ij}^{-}}=(1/\sqrt{2})(\ket{ij}-\ket{ji})$
$(i,j=0,\ldots,m-1,\;i<j)$. Consequently, the general form of
$Q_{A}$ is given by
\begin{eqnarray}\label{general_form}
Q_{A}=\sum_{i<j}\sum_{k<l}\alpha_{ij,kl}\ket{\varphi_{ij}^{-}}\!\bra{\varphi_{kl}^{-}},
\end{eqnarray}
where $\alpha$s must obey $\alpha_{ij,kl}=\alpha_{kl,ij}^{*}$ (to
guarantee Hermiticity) and corresponding positivity conditions.

The general conclusion following the above analysis is that if we
have a decomposable witness $\omega=P+Q^{\Gamma}$ with $P,Q\geq 0$
satisfying the condition $\Trr_B(P+Q)=\mathbbm{1}_{m}/m$ and there
exists a matrix $A$ such that $P_A=0$ and $Q_A$ supported on
$\mathcal{A}_m$, then the corresponding map $\Lambda$ is unital
and detects all the isotropic states implying that its SPA is
entanglement breaking.

Still more can be said about witnesses that satisfy assumption of
the theorem 2. First, let us notice that in general $r(A)\geq 2$.
This follows from the fact that rank-one $A$ lead to $Q_A$ of the
separable form, and such $Q_A$ cannot be supported on
$\mathcal{A}_m$. On the other hand, for full rank $A$ the identity
$P_A=0$ implies $P=0$. Moreover, in such case the relevant $Q$s
are given by
$(A^{-1}\ot\mathbbm{1}_m)\widetilde{Q}(A^{-1}\ot\mathbbm{1}_{m})^{\dagger}$
with $\widetilde{Q}$ of the form (\ref{general_form}) with
$\alpha$s satisfying the following system of equations
\begin{eqnarray}\label{par_cond}
\sum_{\underset{q<i}{i=0}}^{m-1}\alpha_{pi,qi}
+\sum_{\underset{i<p}{i=0}}^{m-1}\alpha_{ip,iq}
-\sum_{\underset{p<i<q}{i=0}}^{m-1}\alpha_{pi,iq}=\frac{2}{m}\bra{p}AA^{\dagger}\ket{q}\nonumber\\
\quad (p,q=0,\ldots,m-1,\;p\leq q).
\end{eqnarray}

In the particular case $A=\mathbbm{1}_{m}$, the above analysis
tells us that all witnesses $W=Q^{\Gamma}$ with $Q$ supported on
$\mathcal{A}_m$, and with maximally mixed first subsystem, satisfy
the conjecture. This particular result is a generalization of the
analysis done for $3\ot 3$ systems in Ref. \cite{Korbicz08PRA}.

Let us now discuss the case of full rank $A$. In the case of
$m=2$, the antisymmetric subspace of two-qubit space is
one-dimensional and spanned by $\ket{\varphi_{01}^{-}}$. To find
$Q$ we need to solve the equation $Q_A=\alpha P_{01}^{-}$ with
$P_{01}^{-}$ denoting projector onto $\ket{\varphi_{01}^{-}}$.
Since $\Trr_B P_{01}^{-}=\mathbbm{1}_{2}/2$, this equation implies
that $AA^{\dagger}=\alpha\mathbbm{1}_{2}$. We know, however, that
$A$ corresponds to normalized pure state and therefore
$\Tr(AA^{\dagger})=2$, which implies that $\alpha=1$ and
consequently $A$ can only be a unitary matrix. Thus, in the
two-qubit case only witnesses being local unitary rotations of
$P_{01}^{-}$ are "detected" by theorem 2.

In the case of $m=3$, $\mathcal{A}_3$ is three-dimensional and
spanned by the vectors $\ket{\varphi_{01}^{-}}$,
$\ket{\varphi_{02}^{-}}$, and $\ket{\varphi_{12}^{-}}$. Every $Q$
defined on $\mathcal{A}_3$ is characterized by nine real
parameters. On the other hand, for any nonsingular $A$,
(\ref{par_cond}) give us exactly nine equations for $\alpha$s,
meaning that generically for any such $A$ there exists a unique
solution to (\ref{par_cond}), and thus a unique witness $\omega$
that satisfies the conjecture. In the particular case of
$A=\mathbbm{1}_{m}$ this witness is given by
$W=(1/3)(P_{01}^{-}+P_{02}^{-}+P_{12}^{-})^{\Gamma}$.

Generically (except for the case $m=2$) the system
(\ref{par_cond}) gives us $m$ equations for $m(m-1)/2$ real
parameters $\alpha_{ij,ij}$ and $m(m-1)/2$ equations for
$m(m-1)[m(m-1)-2]/8$ for complex off-diagonal $\alpha_{ij,kl}$.
Thus, we have in total $m^{2}$ equations for $m^{2}(m-1)^{2}/4$
real parameters. Consequently, for any nonsingular $A$
corresponding to a pure normalized state $\ket{\psi}$, we have a
$[m^{2}(m+1)(m-3)/4]$-parameter class (without taking into account
positivity conditions) of witnesses satisfying the conjecture. It
should be also emphasized that all such witnesses are optimal.
This follows from the fact that since the support of $Q_A$ belongs
to $\mathcal{A}_m$, its kernel contains all the vectors
$(\ket{0}+\alpha\ket{1}+\ldots+\alpha^{m-1}\ket{m-1})^{\ot 2}$
with $\alpha\in\mathbb{C}\cup\{\infty\}$. Then, one knows that the
partial conjugated vectors
$(\ket{0}+\alpha\ket{1}+\ldots+\alpha^{m-1}\ket{m-1})\ot
(\ket{0}+\alpha^{*}\ket{1}+\ldots+(\alpha^{*})^{m-1}\ket{m-1})$
span the whole Hilbert space $\mathcal{H}_m$. Since $Q$ is related
to $Q_A$ via a local nonsingular transformation, $Q^{\Gamma}$ has
mean values on product vectors that also span $\mathcal{H}_m$ and,
according to what was said in Sec. \ref{optimality}, it is
optimal.

In the remaining cases with respect to the rank of $A$, i.e.,
$2\leq r(A)\leq m-1$, characterization of witnesses which are
detected by theorem 2 is a nontrivial task. Still, however, the
above analysis allows us to provide examples of entanglement
witnesses $W=P+Q^{\Gamma}$ obeying the conjecture with are
certainly not optimal. For instance let
$A=\sum_{i=0}^{k-1}\proj{i}$ ($r(A)=k<m$) and
$P=(1/m)\sum_{i=k}^{m-1}\proj{ii}$, and $Q=[2/m(k-1)]\sum_{0\leq
i<j\leq k-1}P_{ij}^{-}$. Then, one easily finds that $P_A=0$ and
$Q_A=Q$.
Finally, $\Trr_BW=\Trr_B(P+Q)=\mathbbm{1}_{m}/m$ and therefore $W$
and the corresponding map obey the conjecture, nevertheless, $W$
is certainly not optimal.

Interestingly, using theorem 1, theorem 2 can be generalized to a
class of maps which contains unital maps as a
special case. More precisely, we have the following statement.\\

\noindent{\bf Theorem 3.} {\it Let $\Lambda:M_m(\mathbb{C})\to
M_m(\mathbb{C})$ be a positive map of the form
$\Lambda=\Lambda_{\mathrm{un}}\circ \Lambda_A$, where
$\Lambda_A(\cdot)=A(\cdot)A^{\dagger}$ and $\Lambda_{\mathrm{un}}$
is a unital map detecting all entangled states in the class
(\ref{class}) with $\ket{\psi}=\mathbbm{1}_m\ot A\smes$. Then SPA
of $\Lambda$ is entanglement breaking.\\}

\noindent{\bf Proof.} The proof goes along the same lines as the
one given in Ref. \cite{Pytel09}. The structural physical
approximation of $\Lambda$ is given by
$\widetilde{\Lambda}[p]=(1-p)D+p\Lambda$. In order to check for
which values of $p$ it is CP, let us notice that the partial
application of $\widetilde{\Lambda}[p]$ to $\mes$ can be expressed
as
\begin{eqnarray}
(I\ot\widetilde{\Lambda}[p])(\mes)&=&\frac{1-p}{m^{2}}\mathbbm{1}_m\ot\mathbbm{1}_m
+p(I\ot\Lambda)(\mes)\nonumber\\
&=&(I\ot\Lambda_{\mathrm{un}})(\rho(p)),
\end{eqnarray}
where to get the second equality we utilized the fact that
$\Lambda=\Lambda_{\mathrm{un}}\ot\Lambda_{A}$ and
$\mathbbm{1}_m\ot A\smes=\ket{\psi}$. Now, we know that $\rho(p)$
is separable for $p\leq 1/(d^{2}\theta+1)$ and therefore
$(I\ot\widetilde{\Lambda}_{p})(\mes)$ is positive and separable
for these values of $p$. On the other hand, by assumption,
$\Lambda_{\mathrm{un}}$ detects all the entangled states $\rho(p)$
meaning that $\widetilde{\Lambda}[p]$ is not CP for
$p>1/(d^{2}\theta+1)$. Consequently, SPA of $\Lambda$ is
entanglement breaking. $\blacksquare$

A particular example of a map satisfying assumptions of the above
fact is $\Lambda=T\circ \Lambda_A$ with $A$ being any matrix. It
is clear that $T$ is unital and detects all the entangled states
$\rho(p)$. On the other hand, one sees that maps of the form
$T\circ \Lambda_A$ do not satisfy assumptions of theorem 2.

\section{Extending the conjecture}
\label{Extending}

It is interesting to ask whether other than $D$ EBCs when used to
construct SPA (according to (\ref{FinSPA})) can also lead to EBCs.
Or, in other words, if all separable states which when put in Eq.
(\ref{SPAWit}) instead of the maximally mixed state, will also
give separable states. Certainly, these can only be the full rank
separable states as otherwise the negative eigenvalues of some
witness $W$ would not be "covered".

It is the purpose of this section to explore the above questions.
We give an example of some other than $D$ entanglement breaking
channel that when used for construction of SPAs does not
necessarily lead to an entanglement breaking map. On the other
hand, it is shown that for any positive map (EW) there exists an
EB channel (separable state) such that the SPA constructed in this
way is also EB (separable). Finally, we comment on the possibility
of extension of the conjecture to the case of Gaussian states.

\subsection{SPAs using other entanglement breaking channels}
\label{Other}

Let $\Phi$ denote some entanglement breaking channel such that the
separable state $\sigma_{\mathrm{sep}}=(I\ot \Phi)(\mes)$ is of
full rank. Then let us consider the following mixture
\begin{equation}\label{constr}
\widetilde{\Lambda}[p]=(1-p)\Phi+p\Lambda,
\end{equation}
where $\Lambda$ is some positive map. For the largest value of
$p$, $p_*$, for which the map $\widetilde{\Lambda}[p]$ is CP we
obtain a completely positive approximation of $\Lambda$. Our aim
now is to check if for all $\Phi$, $\widetilde{\Lambda}[p_*]$ is
EB. Notice, that through the C-J isomorphism the same
approximation as in (\ref{constr}) is defined for EWs.

At the beginning let us consider the cases of two-qubits and
qubit-qutrit. For any optimal witness $W=Q^{\Gamma}$ with $Q\geq
0$ it is then clear that the matrix
\begin{equation}
\widetilde{W}(p)=(1-p)\sigma_{\mathrm{sep}}+pQ^{\Gamma} \qquad
(0\leq p\leq 1)
\end{equation}
represents a separable state for any value of $p$ for which
$\widetilde{W}(p)\geq0$. This is because, as one may easily check,
$[\widetilde{W}(p)]^{\Gamma}$ is positive for any $p$, and
therefore, due to \cite{Horodecki96PLA}, $\widetilde{W}(p)$ is
separable. It is then tempting to ask wether similar results hold
for higher-dimensional systems. In what follows we show that this
not the case. For this purpose we introduce a simple one-parameter
class of entanglement breaking channels corresponding to the known
Werner states \cite{Werner89PRA}. We then show that while the CP
approximations for the the transposition and reduction maps
constructed using these channels are always EB, there exist a
range of the parameter for which CP approximation of the
Breuer-Hall map \cite{Breuer06PRL,Hall06JPA} is no longer an EB
channel.

Let us first introduce a completely positive map
$R_{+}(X)=[1/(m+1)][\Tr(X)\mathbbm{1}_{m}+ X]$. From both maps
$R_{\pm}$ (see Sec. \ref{ClassIn} for the definition of $R_{-}$)
we construct the following channel
\begin{equation}
\Phi[\mu]= T\circ \left[\mu R_{-}+(1-\mu) R_{+} \right]\qquad
(0\leq \mu\leq 1).
\end{equation}
For $\mu=1/2$ the above map reduces to the fully depolarizing
channel, i.e., $\Phi[1/2]=D$. Via the C-J isomorphism, $\Phi[\mu]$
corresponds to the so-called Werner states \cite{Werner89PRA}:
\begin{equation}
\varrho_W(\mu)=\frac{2\mu}{m(m-1)}P_{\mathrm{a}}+\frac{2(1-\mu)}{m(m+1)}P_{\mathrm{s}}\qquad
(0\leq\mu\leq 1),
\end{equation}
where, as previously, $P_{\mathrm{a}}$ and $P_{\mathrm{s}}$ denote
projectors onto antisymmetric and symmetric subspaces of
$\mathcal{H}_m$, respectively. These states are separable iff
$\mu\leq 1/2$ and consequently for this range of $\mu$ the map
$\Phi[\mu]$ represents an EB channel. Moreover, for $\mu\in(0,1)$,
$\varrho_W(\mu)$ is of full rank and therefore $\Phi[\mu]$ is a
good candidate for the construction of completely positive
approximations of positive maps according to (\ref{constr}).

We can now test if such an approximations constructed using the
map $\Phi[\mu]$ give EBCs. Let us start from the transposition
map. From (\ref{constr}) we have
\begin{eqnarray}\label{SPATr}
\widetilde{T}[p,\mu]=(1-p)\Phi[\mu]+pT\qquad (0\leq p\leq 1).
\end{eqnarray}
Partial application of the above map to the maximally entangled
state $P_{m}^{+}$ gives
\begin{equation}\label{SPATr2}
\widetilde{W}_T(p,\mu)=(1-p)\varrho_{W}(\mu)+\frac{p}{m}V_m,
\end{equation}
where $V_m=m[\mes]^{\Gamma}$. It immediately follows (all the
detailed calculations for this and the following cases may be
found in \ref{Appendix}) that $\widetilde{W}_T(p,\mu)\geq 0$ if
and only if
\begin{equation}\label{pTr}
p\leq \frac{2\mu}{2\mu+m-1}.
\end{equation}
Then, it is easy to see that for $p=p_*$, the state
$\widetilde{W}_T(p_*,\mu)$ is separable for $\mu\in(0,1/2]$. Thus,
the approximation of the transposition map (\ref{SPATr}) build
using $\Phi[\mu]$ is also an EBC for any $\mu\in(0,1/2]$.

Now, let us test an approximation (\ref{constr}) for next optimal
decomposable map, the reduction map $R_{-}$. After mixing it with
$\Phi[\mu]$, one gets
$\widetilde{R}_{-}[p,\mu]=(1-p)\Phi[\mu]+pR_{-}$ $(0\leq p\leq
1)$. Via the C-J isomorphism we arrive at the matrix
\begin{equation}\label{red}
\widetilde{W}_{r}(p,\mu)=(1-p)\varrho_{W}(\mu)+\frac{p}{m(m-1)}(\mathbbm{1}_{m}\ot\mathbbm{1}_{m}-m\mes).
\end{equation}
It follows (see \ref{Appendix}) that $\widetilde{W}_{r}(p,\mu)\geq
0$ iff
\begin{equation}\label{pRed}
p\leq \frac{2(1-\mu)}{3+m-2\mu}.
\end{equation}
Then, for the threshold value of $p$, $\widetilde{W}_{r}(p_*,\mu)$
is separable for $\mu\in(0,1/2]$ and therefore again the CP
approximation of $R_{-}$ constructed with the aid of $\Phi[\mu]$
is EB.

So far, we have studied only decomposable maps. Let us now
consider an example of an optimal indecomposable positive map. One
of the commonly known examples is the so-called Breuer-Hall map
\cite{Breuer06PRL,Hall06JPA} (notice that this map was recently
generalized in Refs. \cite{Pytel09,Pytel10}) defined as
\begin{equation}
\Lambda_{U}(X)=\frac{1}{m-2}\left[\Tr(X)\mathbbm{1}_{m}-X-UX^{T}U^{\dagger}\right],
\end{equation}
where now $m$ is even, $U^{T}=-U$, and $U^{\dagger}U=
\mathbbm{1}_{m}$.

Partial application of the approximation of $\Lambda_U$,
$\widetilde{\Lambda}_{U}[p,\mu]=(1-p)\Phi[\mu]+p\Lambda_U$ to
$P_{m}^{+}$ gives
\begin{eqnarray}
&&\fl\widetilde{\mathcal{W}}(p,\mu)
=\frac{1}{m}\left[\left(\frac{2\mu(1-p)}{m-1}
+\frac{p}{m-2}\right)P_{\mathrm{a}}\right.
\left.+\left(\frac{2(1-\mu)(1-p)}{m+1}+\frac{p}{m-2}\right)P_{\mathrm{s}}\right]\nonumber\\
&&\hspace{-0.9cm}-\frac{p}{m-2}\mes-\frac{p}{m(m-2)}V_{U},
\end{eqnarray}
where $V_{U}=(\mathbbm{1}_{m}\ot U)V(\mathbbm{1}_{m}\ot
U^{\dagger})$. One sees that $\widetilde{\mathcal{W}}(p,\mu)\geq
0$ iff
\begin{equation}\label{pBH}
p\leq \frac{2(1-\mu)}{2(1-\mu)+m+1}.
\end{equation}
It then follows that for $p=p_*$ the partial transposition of
$\widetilde{\mathcal{W}}(p_*,\mu)$ is not positive whenever
$\mu<[(m-1)(m-2)]/[2(m^{2}-2)]$, meaning that for $m\geq 3$,
$\widetilde{\mathcal{W}}(p_*,\mu)$ is not always separable and
therefore at least for the above values of $\mu$ the CP
approximation (\ref{constr}) of $\Lambda_U$ is certainly not EB.
As a result the conjecture does not hold for any EBC $\Phi$ such
that $(I\ot \Phi )(P_m^{+})$ is of full rank.


\subsection{Relaxing the conjecture}

As shown in the preceding section, the conjecture does not have to
hold when instead of $D$, one uses some other EB channel $\Phi$
such that $(I\ot \Phi)(P_m^{+})$ is of full rank. Nevertheless,
the above examples show that at least in the case of the
transposition and reduction maps, the approximations constructed
from the channel $\Phi[\mu]$ give EB map. It is then tempting to
conjecture that for any positive map $\Lambda$ there exists at
least one EB channel such that the approximation of $\Lambda$ with
$D$ substituted by this EB channel is again EB. Here we show that
this statement is true for any positive $\Lambda$.

In order to proceed more formally let us consider a pure entangled
state $\ket{\psi}\in\mathcal{H}_{mn}$ and the set of states
\begin{equation}\label{sep}
S_{\psi}=\{\varrho\in\mathcal{D}_{m,n}|\langle\psi|\varrho|\psi\rangle=0\}.
\end{equation}
In general, the set $S_{\psi}$ may contain all types of states
with respect to separability, i.e., NPT and PPT entangled as well
as separable states. Let us show that the latter constitute a
subset of $S_{\psi}$ with nonempty interior in $S_{\psi}$.\\

\noindent {\bf Lemma.} {\it The subset of separable states in
$S_{\psi}$ has nonempty interior in $S_{\psi}$.\\}

\noindent {\bf Proof.} For the sake of simplicity and clearness we
assume that $m=n$ and that Schmidt rank of $\ket{\psi}$ is $m$.
Then, since $\ket{\psi}$ is related to $\ket{\psi_{m}^{+}}$ via
$\ket{\psi}=A\ot \mathbbm{1}_m\ket{\psi_{m}^{+}}$ with full rank
$A$, it suffices prove the above statement for
$\ket{\psi}=\ket{\psi_{m}^{+}}$. It is nevertheless clear that the
proof is also valid when $m\neq n$ and the Schmidt rank of
$\ket{\psi}$ is less than $\min\{m,n\}$.

We will show, following the proof of theorem 1 from Ref.
\cite{Zyczkowski98PRA}, that not all the separable states from
$S\equiv S_{\psi_{m}^{+}}$ can be approached arbitrarily close by
entangled states from $S$. First, however, let us make some
statements about the subset of pure separable states in $S$,
denoted henceforward by $S_{\mathrm{sep}}$. It follows from
(\ref{sep}) that they satisfy the relation $\langle
e,f|\psi_{m}^{+}\rangle=0$, which after simple calculations,
implies that elements of $S_{\mathrm{sep}}$ are those product
vectors for which $\ket{f}\in\mathbbm{C}^{m}$ is any vector
orthogonal to $\ket{e^{*}}$. It is clear that they span
$(m^{2}-1)$-dimensional subspace in $\mathcal{H}_m$, denoted by
$\widetilde{\mathcal{H}}$, while projectors onto these vectors
span $(m^2-1)^{2}$-dimensional subspace in the Hilbert space of
$m^{2}\times m^{2}$ matrices with the Hilbert-Schmidt scalar
product, denoted $\mathcal{V}_\mathrm{sep}$.

We are now ready to proceed with the main part of the proof. Let
us take a separable state $\sigma_{\mathrm{sep}}\in S_{\psi}$
being a continuous convex combination over all product states from
$S_{\mathrm{sep}}$, i.e.,
\begin{equation}
\sigma_{\mathrm{sep}}= \sum_{\ket{e,f}\in S_{\mathrm{sep}}}
p(e,f)\proj{e,f},\quad \sum_{\ket{e,f}\in S_{\mathrm{sep}}}
p(e,f)=1,
\end{equation}
where $p(e,f)>0$ for any $\ket{e,f}\in S_{\mathrm{sep}}$, and
assume that there exists a sequence of entangled states
$\varrho_{n}\in S_{\psi}$ convergent to $\sigma_{\mathrm{sep}}$ in
the Hilbert-Schmidt norm. Then, it follows from Ref.
\cite{Horodecki96PLA} that for any $\varrho_n$ there exists an EW
$W_n$ detecting it, i.e., such that $\Tr(W_n\varrho_n)<0$.

We can obviously write $W_{n}=O_n+O_n^{\perp}$, where $O_n\in
\mathcal{V}_{\mathrm{sep}}$ and $O_n^{\perp}$ belongs to the
subspace of $M_{m,m}$ orthogonal to $\mathcal{V}_{\mathrm{sep}}$
and spanned by $P_m^{+}$, $\ket{\phi}\!\bra{\psi_m^{+}}$, and
$\ket{\psi_m^{+}}\!\bra{\phi}$ with arbitrary $\ket{\phi}\in
\widetilde{\mathcal{H}}$. Since
$\ket{\psi_{m}^{+}}\in\mathrm{ker}(\varrho_{n})$, it is clear that
for any $n$, $\Tr(W_n\varrho_n)=\Tr(O_n\varrho_{n})< 0$ and hence
$O_n$ are certainly nonzero matrices. Consequently, we can
introduce normalized matrices
$\widetilde{O}_{n}=O_{n}/\|O_{n}\|_2$ which now belong to the
compact set of matrices from the unit sphere in $M_{m,m}$. As a
result, there exists a subsequence $\widetilde{O}_{n_{k}}$
converging to some element
$\widetilde{O}\in\mathcal{V}_{\mathrm{sep}}$ with
$\|\widetilde{O}\|_2=1$. Then, analogous estimation to the one in
\cite{Zyczkowski98PRA} gives the following chain of inequalities
\begin{equation}
0\leq \Tr(\widetilde{O}_{n_{k}}\sigma_{\mathrm{sep}})\leq
\left\|\sigma_{\mathrm{sep}}-\varrho_{n_{k}}\right\|_{2},
\end{equation}
which immediately implies that
\begin{equation}\label{relation}
\Tr(\widetilde{O}\sigma_{\mathrm{sep}})=0.
\end{equation}

On the other hand, $\langle e,f|W_n|e,f\rangle=\langle
e,f|O_n|e,f\rangle\geq 0$ for any $\ket{e,f}\in S_{\mathrm{sep}}$,
which, due to the continuity of the scalar product, yields
\begin{equation}
\langle e,f|\widetilde{O}|e,f\rangle\geq 0\qquad (\ket{e,f}\in
S_{\mathrm{sep}}).
\end{equation}
The latter together with (\ref{relation}) imply that $\langle
e,f|\widetilde{O}|e,f\rangle=0$ for any $\ket{e,f}\in
S_{\mathrm{sep}}$, which, due to the fact that
$\widetilde{O}\in\mathcal{V}_{\mathrm{sep}}$, allows to infer that
$\widetilde{O}$ is a zero matrix. This, however, leads to a
contradiction with the fact that $\|\widetilde{O}\|_2=1$.
$\blacksquare$

Now we are prepared to proceed with our theorem.\\

\noindent {\bf Theorem 4.} {\it Let $W$ be some EW acting on
$\mathcal{H}_{m,n}$. Then there exists a separable state
$\Sigma_{\mathrm{sep}}$ such that the approximation of $W$
constructed using $\Sigma_{\mathrm{sep}}$ according to
(\ref{constr}) is separable. In other words, for any $\Lambda$
there exists an EBC $\Phi$ such that the CP approximation of
$\Lambda$ constructed using $\Phi$ is EB.\\}

\noindent {\bf Proof.} First let us construct the standard SPA of
$W$. Denoting by $-\lambda$ and by $\ket{\psi}$ the smallest
eigenvalue of $W$ and the corresponding eigenvector, respectively,
its SPA reads $\widetilde{W}=[1/(1+\lambda
mn)](\lambda\mathbbm{1}_{mn}+W)$. Let us assume that the latter is
entangled as otherwise there is nothing to prove.

Clearly $\langle\psi|\widetilde{W}|\psi\rangle=0$ and therefore
the SPA of $W$ brings us to the set $S_{\psi}$ (see (\ref{sep})).
Naturally, now we can use the above lemma. Precisely, we can take
a separable state $\sigma_{\mathrm{sep}}$ from the interior of the
set of separable states in $S_{\psi}$, and mix it with
$\widetilde{W}$ so that the resulting density matrix is separable.
Since $\ket{\psi}\in\mathrm{ker}(\sigma_{\mathrm{sep}})$,
$\sigma_{\mathrm{sep}}$ does not cover the lowest eigenvalue of
$W$. Therefore we can treat the maximally mixed state used
previously to construct $\widetilde{W}$ together with
$\sigma_{\mathrm{sep}}$ as a postulated separable state allowing
to build the approximation of $W$.

Let us now proceed more formally. Due to the very definition of
$\sigma_{\mathrm{sep}}$, there exists $q\in(0,1)$ such that
$\varrho(q)=q\sigma_{\mathrm{sep}}+(1-q)\widetilde{W}$ is a
separable state. Let $q_{*}$ denote a minimal value of the mixing
parameter for which $\varrho(q)$ is separable. Then, one can check
that the state we can use to construct the nonstandard SPA of $W$
is
\begin{equation}
\Sigma_{\mathrm{sep}}=\frac{1-q_{*}}{q_*+\lambda
mn}\left(\lambda\mathbbm{1}_m\ot\mathbbm{1}_n +\frac{1+\lambda
mn}{1-q_{*}}q_{*}\sigma_{\mathrm{sep}}\right).
\end{equation}
For this purpose, let us consider the mixture
$\overline{W}(r)=(1-r)\Sigma_{\mathrm{sep}}+rW$. From the facts
that the lowest eigenvalue of $W$ is $-\lambda$ and
$\langle\psi|\sigma_{\mathrm{sep}}|\psi\rangle=0$, we see that
$\overline{W}(r)$ is positive for $r\leq (1-q_*)/(1+\lambda mn)$.
It remains to show that for the threshold value $r_*=
(1-q_*)/(1+\lambda mn)$, $\overline{W}(r_{*})$ is separable.
Direct check shows that $\overline{W}(r_{*})$ is exactly
$\varrho(q_{*})$, which we know to be separable. This finishes the
proof. $\blacksquare$

Let us now comment on the obtained result. It is clear that if the
set $S_{\psi}$ consisted only of separable states, the conjecture
would be immediately proven. However, in general this is not the
case. Consequently, one way to prove it is to show that mixing
given witness with maximally mixed state we always "land" in the
set of separable states belonging to $S_{\psi}$.

\subsection{The conjecture in continuous-variable systems}

In the present section we propose an extension of the notion of
structural physical approximation onto the case of Gaussian
positive maps, and we study the conjecture for the transposition
map. First, however, we need to remind some basic concepts and
notions from the theory of Gaussian states, operators and maps.

The Hilbert space describing $n$ harmonic oscillators ($n$ modes)
is $\mathcal{H}=L^2(\mathbb{R}^{n})$. To the $i$th mode there are
associated two canonical observables $X_i$ and $P_i$
($i=1,\ldots,n$). All these operators fulfil the canonical
commutation relations $[X_k,P_l]=\mathrm{i} \delta_{kl}$. Every
operator $A$ acting on $B(\mathcal{H})$ can be represented in the
following form \cite{Hayashi07PR,intquantopt}:
\begin{equation}
A=\pi^{-n}\int_{R^{2n}} d^{2n}\xi\, \chi_A(\xi)W_{-\xi},
\end{equation}
where $W_\xi=\exp(-\mathrm{i}\xi^T R)$ is the so-called Weyl
operator (displacement operator), $R = (X_1,P_1,\ldots,X_n,P_n)$
and $\chi_A(\xi)=\Tr(A W_\xi)$ is the characteristic function of
$A$. An operator $A$, whose characteristic function $\chi_A(\xi)$
has the following form
\begin{eqnarray}
\chi_A(\xi)= \exp\left( -\frac{1}{4}\xi^T \gamma \xi + \mathrm{i}
d^T \xi - c\right), \label{gaussstate}
\end{eqnarray}
is called {\it a Gaussian operator} \cite{hpaqt}. In the above
$\gamma$ and $d$ denote a covariance-like matrix and a
displacement-like vector with entries given by
\begin{equation}
\gamma_{ij}=2\mathrm{Re}\{\Trr[(R_{i}-d_{i})(R_{j}-d_{j})A]\},\qquad
d_{i}=\Trr(R_{i}A),
\end{equation}
and $c$ stands for the normalization constant. Let the set of all
Gaussian operators acting on $\mathcal{H}$ be denoted by
$Q(\mathcal{H})$. Among all the elements of $Q(\mathcal{H})$ we
distinguish those for which $\gamma, d$, and $c$ are real and
$\gamma$ obeys $\gamma+\mathrm{i} J_n\geq 0$ with $J_{n}$ defined
as
\begin{equation}
J_n=\bigoplus_{i=1}^n J,\qquad J = \left(
\begin{array}{cc}
0&1\\
-1&0\\
\end{array}
\right),
\end{equation}
These elements are just Gaussian states, and the corresponding
$\gamma$s are proper covariance matrices.

A linear map $\Lambda:B(\mathcal{H})\mapsto B(\mathcal{H'})$ that
maps Gaussian operators to Gaussian operators, i.e.,
$\Lambda[Q(\mathcal{H})]\subseteq Q(\mathcal{H'})$ is called {\it
Gaussian}. In a full analogy to the finite-dimensional case, one
defines positive, completely positive (among them also quantum
channels), and entanglement breaking Gaussian maps \cite{Holevo}.
Let us only remind \cite{EisertWolf} that the action of any {\it
Gaussian channel} $\mathcal{G}:B(\mathcal{H})\mapsto
B(\mathcal{H})$ can be represented on the ground of covariance
matrices as
\begin{eqnarray}\label{GaussianCH}
\gamma'= X^T \gamma X + Y,\qquad d'=X^T d + v,
\end{eqnarray}
where $\gamma'$ ($d'$) and $\gamma$ ($d$) are covariance matrices
(displacement vectors) of the Gaussian states after and before an
application of $\mathcal{G}$, respectively. Then, $X$ and $Y$ are
some real $2n\times 2n$ matrices, which in order to guarantee
complete positivity of $\mathcal{G}$ must obey
\begin{eqnarray} \label{cppcond}
&Y + \mathrm{i} J_n - \mathrm{i} X^T J_n X\geq 0.&
\end{eqnarray}
Then, it was shown in Ref. \cite{Holevo} that a given Gaussian
channel is entanglement breaking if and only if $Y$ can be
decomposed as $Y=A+B$ with real and symmetric $A$ and $B$
satisfying
\begin{eqnarray}\label{ebcond}
A\geq -\mathrm{i} J_n , \qquad B\geq -\mathrm{i} X^T J_n X.
\end{eqnarray}

Having recalled all the necessary notions we can pass to the
concept of SPA in Gaussian maps. To the best of our knowledge, the
first attempt to get the best approximation (in terms of the
fidelity) of the transposition map in continuous-variable systems
was made in Ref. \cite{Buscemi03PLA}. The authors looked for the
best completely positive approximation (in terms of the fidelity)
of the transposition map under the assumption that it has to work
equally well for all pure states. As this cannot be achieved in
the infinite-dimensional case, they proposed approximation which
depends on the input states.

Here we follow a different approach, which is more consistent with
the standard SPA (\ref{FinSPA}) in the sense that it does not
dependent on the input state. In order to proceed more formally
let $\Lambda:B(\mathcal{H})\mapsto B(\mathcal{H})$ be some
positive Gaussian map. We know from Ref. \cite{giedkecirac} that
on the level of CMs its action can be represented by some map
which for simplicity we will also denote by $\Lambda$. Clearly, we
cannot construct SPA as in Eq. (\ref{FinSPA}) as addition of an
identity to the gaussian density operator may easily throw us out
of the set of density operators. Nevertheless, as we will see
below, the above reasoning can successfully be adopted, at least
in some cases, on the level of covariance matrices. This is
because an addition of $\mathbbm{1}$ with some weight $p$ to the
covariance matrix $\gamma'=\Lambda(\gamma)$, i.e.,
\begin{eqnarray}
&\widetilde{\Lambda}_p(\gamma)= \Lambda(\gamma) + p \mathbbm{1},
\qquad \widetilde{\Lambda}_p(d)= \Lambda(d), \label{spadef}&
\end{eqnarray}
keeps us in the set of proper covariance matrices. On the level of
density operators the above means composition of $\Lambda$ with
Gaussian channel $\Phi[p]$ defined as
\begin{equation}
\Phi_p(\rho) =\frac{1}{\pi^n p^n}\int \mathrm{d}^{2n}\xi\,
W_\xi\rho W^\dagger_\xi \mathrm{e}^{-\frac{\xi^T\xi}{ p} }.
\end{equation}
From the physical point of view, the latter adds classical
gaussian noise with variance $p$ to a given state
\cite{Hayashi07PR}. Consequently, on the level of density
matrices, our SPA is defined as
$\widetilde{\Lambda}[p]=\Phi[p]\circ\Lambda$.

Now, we see that if the positive Gaussian map $\Lambda$ can be
represented as in Eq. (\ref{GaussianCH}), one can always find $p$
such that $Y + p \mathbbm{1}+ \mathrm{i} J_n - \mathrm{i} X^T J_n
X\geq 0 $. Consequently, at least for any positive Gaussian map
which action on covariance matrices is given by Eq.
(\ref{GaussianCH}), the approximation defined by Eq.
(\ref{spadef}) leads to a completely positive map.

Let us now consider the conjecture in the context of the
transposition map. We will show that the Gaussian SPA of
transposition map leads to the entanglement breaking channel. To
this aim recall that the action of transposition map on an
$n$-mode Gaussian state (characterized by $\gamma$ and $d$) takes
form \cite{Simon00PRL}:
\begin{eqnarray}\label{spaofscm}
\gamma'=\Delta \gamma \Delta,\qquad d'=\Delta d,
\end{eqnarray}
where
\begin{eqnarray}
\Delta = \bigotimes_{i=1}^n \left(
\begin{array}{cc}
1&0\\
0&-1
\end{array}
\right).
\end{eqnarray}
From Eqs. (\ref{spadef}) and (\ref{spaofscm}) one infers that the
action of SPA of the transposition map
$\widetilde{T}[p]=\Phi[p]\circ T$ on the level of covariance
matrices reads $\widetilde{T}[p](\gamma)=T(\gamma)+p\mathbbm{1}$.
The condition (\ref{cppcond}) tells us that $\widetilde{T}[p]$ is
completely positive iff
\begin{eqnarray}
p \mathbbm{1} + \mathrm{i} \bigotimes_{i=1}^n \left(
\begin{array}{cc}
0&2\\
-2&0
\end{array}
\right) \geq 0, \label{cppsc}
\end{eqnarray}
which is satisfied for $p\geq 2$. Let us now check if for the
threshold value $p_*=2$, the map $\widetilde{T}[2]$ is
entanglement breaking. Indeed, one sees that in this case $Y=2
\mathbbm{1}$ can be decomposed as $Y=A+B$ with $A=B=\mathbbm{1}$
and both matrices $A$ and $B$ fulfill the conditions
(\ref{ebcond}).

Concluding, let us mention that the obtained channel in the
one-mode case can be represented as
\begin{equation}
\widetilde{T}_1(\rho)=\frac{1}{\pi}\int \mathrm{d}^2 \xi \langle
\xi_x,\xi_p|\rho|\xi_x,\xi_p\rangle |\xi_x,-\xi_p\rangle\langle
\xi_x,-\xi_p |,
\end{equation}
where $|\xi_x,\xi_p\rangle$ is a coherent state with the
displacements given by $\xi_x$ and $\xi_p$. In the general case of
$n$ modes we have $T_n=\otimes_{i=1}^n T_1$.

\section{Conlusions}
\label{Concl}

The structural physical approximation
\cite{HorodeckiEkert02PRL,MouraAlves03PRA,PHorodecki03PRA} is one
of the known solutions to the problem of applicability of positive
maps in an experimental detection of entanglement. The recent
observations \cite{Fiurasek02PRA,Korbicz08PRA,Pytel09,Pytel10}
that SPAs of many known optimal positive maps give EB channels or
maps in general, makes this notion interesting also from the
mathematical point of view. Certainly, a full proof of this
conjecture would shed new light on the geometry of convex sets of
states with positive partial transpose and their relations to
entanglement witnesses. Motivated by its mathematical importance,
we have addressed several issues related to the conjecture.

Let us shortly summarize the obtained results. First, we have
extended the set of optimal witnesses satisfying the conjecture.
For instance, we have proven it for witnesses
$W=\proj{\psi}^{\Gamma}$ for any entangled $\ket{\psi}$. Also,
utilizing recent results of Ref. \cite{Acin,Pytel09}, we have
determined the whole class of decomposable witnesses satisfying
the conjecture which as a particular case contain all witnesses
$W=Q^{\Gamma}$ with $Q$ acting on the antisymmetric subspace of
$\mathbbm{C}^{m}\ot\mathbbm{C}^{m}$ with the first subsystem in
the maximally mixed state $\mathbbm{1}_{m}/m$.

Then, we have considered SPAs constructed from other than the
completely depolarizing channel $D$ entanglement breaking
channels. We have shown that there exist EBC which certainly does
not satisfy the statement of the conjecture (see Sec.
\ref{SPAConj}). In other words, there exists EB channels that can
be used to construct SPAs, nevertheless, the obtained CP maps do
not have to be EB. On the other hand, we have proven an
interesting fact that for any positive map $\Lambda$ there exists
an EB map $\Phi$ such that the SPA of $\Lambda$ constructed with
the help of $\Phi$ is EB. Finally, we have asked a similar
question in the case of continuous-variable systems. Precisely, we
have proposed a generalization of the notion of SPA to the
Gaussian case and proven that such SPA of the transposition map is
entanglement breaking.

Let us just shortly sketch the possible lines of further
investigations related to this subject. First of all, the complete
proof of the conjecture is still missing. Then, studies on the
relations of SPAs to the conjecture in the case of multipartite
and continuous-variable systems seem interesting. We only touch
this problem in CV systems and further works clarifying it are
desirable. One knows, however, that the theory of positive maps
serving for detection of entanglement in much better developed in
finite-dimensional systems. This is certainly because the set of
covariance matrices is convex and closed and therefore one can
define entanglement witnesses on the level of covariance matrices
instead of density operators (see e.g. \cite{HyllusEisert}). Also,
the notion of optimality of positive maps in this case is missing.
Finally, following out results, one could investigate if other
SPAs (than the one constructed from $D$) that lead to the
entanglement breaking channels are maybe easier to implement
experimentally.

\setcounter{equation}{0}

\section*{Acknowledgments}
Discussions with A. Ac\'in, J. Stasi\'nska, J. de Vicente, A.
Winter are acknowledged. \L{}. Czekaj gratefully acknowledges ICFO
for kind hospitality. This work was supported by EU IP AQUTE,
Spanish MINCIN project FIS2008-00784 (TOQATA), Consolider Ingenio
2010 QOIT, EU STREP project NAMEQUAM, the Alexander von Humboldt
Foundation, Korea Research Foundation Grant(KRF-2008-313-C00185),
and the Polish Ministry of Science and Higher Education under the
grant No. NN202231937.

\appendix
\section{Detailed calculations for Sec. \ref{Other}}
\label{Appendix}

{\it Transposition map.} Due to the fact that
$V_m=P_{\mathrm{s}}-P_{\mathrm{a}}$, we can rewrite (\ref{SPATr2})
in the following form
\begin{eqnarray}
\fl\widetilde{W}_{T}(p,\mu)&=&\frac{1}{m}\left\{\left[\frac{2\mu(1-p)}{m-1}-p\right]P_{\mathrm{a}}
+\left[\frac{2\mu(1-p)(1-\mu)}{m+1}+p\right]P_{\mathrm{s}}\right\}.
\end{eqnarray}
Since $P_{\mathrm{s}}P_{\mathrm{a}}=0$ it is clear that
$\widetilde{W}_{T}(p,\mu)\geq 0$ if and only if the expression
appearing in the first square bracket is nonnegative. This means
that $\widetilde{T}[p,\mu]$ is completely positive for $p$ obeying
(\ref{pTr}). For the threshold value of $p$, i.e.,
$p_{*}=2\mu/(2\mu+m-1)$, the matrix $\widetilde{W}(p_{*},\mu)$ is
just the normalized projector onto $\mathcal{S}_m$ and therefore
separable.

{\it Reduction map.} Utilizing the fact that
$P_{\mathrm{s}}+P_{\mathrm{a}}=\mathbbm{1}_{m}$ one can rewrite
Eq. (\ref{red}) as
\begin{eqnarray}
\fl\widetilde{W}_{r}(p,\mu)=\frac{1}{m}\left\{\left[\frac{2\mu(1-p)+p}{m-1}\right]P_{\mathrm{a}}
+\left[\frac{2(1-\mu)(1-p)}{m+1}+\frac{p}{m-1}\right]P_{\mathrm{s}}\right\}-\frac{p}{m-1}\mes.\nonumber\\
\end{eqnarray}
Since $\ket{\psi_{m}^{+}}\in \mathcal{S}_m$, $W(p,\mu)\geq 0$ iff
$p$ satisfies (\ref{pRed}). For the threshold value of $p$,
$\widetilde{W}_{r}(p_*,\mu)$ takes the following form
\begin{equation}
\widetilde{W}_{r}(p_*,\mu)=N_1
\left[\frac{1+m\mu}{m}P_{\mathrm{as}}+(1-\mu)(P_{\mathrm{sym}}-\mes)\right]
\end{equation}
with $N_1=2/[(m-1)(3+m-2\mu)]$. In the above one recognizes the
so--called $OO$--invariant states, that is, density matrices which
are invariant under bilateral application of the orthogonal group
$O(m)$. These states were investigated in
\cite{VollbrechtWerner01PRA} and it was shown there that all
$OO$--invariant states with positive partial transposition are
separable. It then suffices to check then spectrum of the partial
transpose of $[\widetilde{W}_{r}(p_*,\mu)]$. After some simple
movements, the latter can be written as
\begin{eqnarray}
\fl[\widetilde{W}_{r}(p_*,\mu)]^{\Gamma}=
N_2\left[(m-1+2\mu)P_{\mathrm{s}}+(m+3-2\mu)P_{\mathrm{a}}\right.
\left.+(m-2\mu m-1)m\mes\right],
\end{eqnarray}
where $N_2=1/[m(m-1)(m+3-2\mu)]$. This implies that
$[\widetilde{W}_{r}(p_*,\mu)]^{\Gamma}\geq 0$ and therefore
according to the results from \cite{VollbrechtWerner01PRA},
$\widetilde{W}_{r}(p_*,\mu)$ is separable for any $\mu\in(0,1/2]$.

{\it The Breuer-Hall map.} In order to determine the range of $p$
for which $\widetilde{\mathcal{W}}(p,\mu)$ is positive we need to
make the following observations. Denoting by
$\ket{\psi_{\mathrm{s}}}$ $(\ket{\psi_{\mathrm{a}}})$ an arbitrary
state belonging to $\mathcal{S}_m$ $(\mathcal{A}_m)$, we see that
$V_{U}\ket{\psi_{\mathrm{s}}}=-U\ot U\ket{\psi_{\mathrm{s}}}$
($V_{U}\ket{\psi_{\mathrm{a}}}=U\ot U\ket{\psi_{\mathrm{a}}}$) for
any real unitary satisfying $U^{T}=-U$. This implies that
$V_U\mathcal{S}_m\subseteq \mathcal{S}_m$ and similarly for
$\mathcal{A}_m$.
It then follows that the smallest eigenvalue of
$\widetilde{\mathcal{W}}(p,\mu)$ corresponds to $\smes$ and
$\widetilde{\mathcal{W}}(p,\mu)$ is positive iff $p$ obeys
(\ref{pBH}). The partial transposition of
$\widetilde{\mathcal{W}}(p_*,\mu)$ reads
\begin{eqnarray}
&&\fl[\widetilde{\mathcal{W}}(p_*,\mu)]^{\Gamma}=N_3\left[\left(m-\frac{2\mu}{m-1}\right)\mathbbm{1}_{m^{2}}-2(1-\mu)V_{m}
\left(
m-2-\frac{2m\mu(m-2)}{m-1}\right)m\mes\right.\nonumber\\
&&\hspace{0.3cm}\left.-2(1-\mu)m(\mathbbm{1}_{m}\ot
U)\mes(\mathbbm{1}_{m}\ot U^{\dagger})\right],
\end{eqnarray}
where $N_3$ is a constant irrelevant for further consideration.
Since $U=-U^{T}$, it follows that $\Trr(U)=0$ and therefore
$\smes$ and $\mathbbm{1}_{m}\ot U\smes$ are orthogonal.
Consequently, for
\begin{equation}
\mu<\frac{(m-1)(m-2)}{2(m^{2}-2)},
\end{equation}
$[\widetilde{\mathcal{W}}(p_*,\mu)]^{\Gamma}$ is not positive and
hence $\widetilde{\mathcal{W}}(p_*,\mu)$ cannot be separable.

\end{document}